\documentclass[twocolumn]{aastex631}

\usepackage[normalem]{ulem}

\graphicspath{{./}{figures/}}

\def\Halpha{\mbox{H\hspace{0.1ex}$\alpha$}\,}
\def\kms{\hbox{km$\;$s$^{-1}$}}

\def\deg{\hbox{$^\circ$}}

\def\HeIIRline{\ion{He}{1}~10830\,\AA\,}

\def\CaIR{\ion{Ca}{2}~8542\,\AA\,}
\def\CaIRII{\ion{Ca}{2}~8662\,\AA\,}

\def\logtau{$\log\tau_{\mathrm{500}}$\,}

\def\blos{$B_{\mathrm{LOS}}$\,}

\def\temp{$T$\,}
\def\vlos{$V_{\mathrm{LOS}}$\,}
\def\vturb{$V_{\mathrm{turb}}$\,}
\def\utw{\ensuremath{\smash{\rlap{\lower5pt\hbox{$\sim$}}}}}
\def\udtw{\ensuremath{\smash{\rlap{\lower6pt\hbox{$\approx$}}}}}

\def\deltab{$\Delta \lambda_{\mathrm{B}}$\,}

\def\FeICaIRIIline{\ion{Fe}{1}~8661.8991~\AA\,}

\begin{document}


\title{Simultaneous spectropolarimetric observations in the \Halpha{} and \CaIRII{} lines of an active region} 

\author[0000-0001-5253-4213]{Harsh Mathur}
\affiliation{Indian Institute of Astrophysics, II Block, Koramangala, Bengaluru 560 034, India}

\author[0000-0002-0465-8032]{K. Nagaraju}
\affiliation{Indian Institute of Astrophysics, II Block, Koramangala, Bengaluru 560 034, India}

\author[0000-0003-4065-0078]{Rahul Yadav}
\affiliation{Laboratory for Atmospheric and Space Physics, University of Colorado, Boulder, CO 80303, USA}

\author[0000-0003-0585-7030]{Jayant Joshi}
\affiliation{Indian Institute of Astrophysics, II Block, Koramangala, Bengaluru 560 034, India}

\begin{abstract}
We present spectropolarimetric observations of an active region recorded simultaneously in the \Halpha{} and the \CaIRII{} lines.
The sunspot exhibits multiple structures, including a lightbridge and a region where \CaIRII{} line core is in emission.
Correspondingly, the \Halpha{} line core image displays brightening in the emission region, with the spectral profiles showing elevated line cores.
The stratification of the line-of-sight magnetic field is inferred through non-LTE multiline inversions of the \CaIRII{} line and the weak field approximation over the \Halpha{} line.
The field strength inferred from the \Halpha{} line core is consistently smaller than that inferred from inversions at \logtau{} = $-$4.5.
However, the study finds no correlation between the WFA over the core of the \Halpha{} line and that inferred from inversions at \logtau{} = $-$4.5.
In regions exhibiting emission features, the morphology of the magnetic field at \logtau{} = $-$4.5 resembles that at \logtau{} = $-$1, with slightly higher or comparable field strengths.
The magnetic field morphology inferred from the core of the \Halpha{} line is also similar to that inferred from the full spectral range of the \Halpha{} line in the emission region.
The field strength inferred in the lightbridge at \logtau{}= $-$1 is smaller than the surrounding umbral regions and comparable at \logtau{} = $-$4.5.
Similarly, the field strength inferred in the lightbridge from the WFA over the \Halpha{} line appears lower compared to the surrounding umbral regions.

\end{abstract}

\keywords{sun:photosphere, chromosphere, magnetic field}

\label{sect:intro}
\section{Introduction}

Simultaneous multiline spectropolarimetry is a powerful observational technique that allows the inference of the magnetic field at multiple heights of the solar atmosphere.

In the recent past, many authors have used spectropolarimetric observations of \CaIR{} / \HeIIRline{} lines recorded simultaneously with lines of \ion{Fe}{1} atom to infer the stratification of the line-of-sight (LOS) magnetic field in various solar features.
Just to mention a few recent studies, the 3-D structure of the magnetic field of sunspots from photosphere to middle chromosphere has been studied by \citet{2019ApJ...873..126M}, \citet{2017A&A...604A..98J}, \citet{2016A&A...596A...8J}, and \citet{2015SoPh..290.1607S}.
Magnetic field variation in umbral flashes has been studied by many authors in recent literature \citep[for example][]{2023A&A...676A..77F, 2023ApJ...945L..27F, 2019A&A...627A..46B, 2018A&A...619A..63J, 2018ApJ...860...28H}.
Magnetic topologies of inverse Evershed flow have also been reported in literature \citep{2022A&A...662A..25P, 2019ApJ...874....6B}.
Changes in the chromospheric field during flares are studied by many authors \citep[][]{2023ApJ...954..185F,  2021A&A...645A...1V, 2021A&A...649A.106Y, 2019A&A...621A..35L, 2018ApJ...860...10K, 2017ApJ...834...26K}.
For a detailed overview of advancements in measurement techniques and analysis methods, we refer the reader to \citet{2017SSRv..210...37L}.

Both \HeIIRline{} and \CaIR{} lines present valuable tools for probing the chromospheric magnetic field due to their relatively well-understood line formation and interpretability using non-local thermodynamic equilibrium (non-LTE) inversion codes \citep[e.g.][]{2000ApJ...530..977S, 2008ApJ...683..542A, 2016ApJ...830L..30D, 2019A&A...623A..74D, 2022A&A...660A..37R}.
However, it is crucial to acknowledge their inherent limitations. 
The formation of the \HeIIRline{} line occurs within a limited range of heights in the upper chromosphere, and its formation relies on incoming EUV radiation from the coronal and transition region \citep{1997ApJ...489..375A, 2016A&A...594A.104L}.
This offers sensitivity to specific atmospheric layers but limits its applicability across diverse solar features.
In contrast, the \CaIR{} line forms from the upper photosphere to the mid-chromosphere, providing broader spatial coverage.
However, in flaring active regions, the \ion{Ca}{2} ion gets ionized to the \ion{Ca}{3} ion, potentially causing the \CaIR{} line to sample deeper layers of the solar atmosphere \citep[][]{2016ApJ...827..101K, 2018ApJ...860...10K, 2019A&A...631A..33B}.
Thus it is challenging to interpret magnetic field measurements in such dynamic environments using the \CaIR{} line.

The spectropolarimetric observations of the \Halpha{} line recorded simultaneously with observations of established spectral lines, such as those of the \ion{Ca}{2} atom can serve as an effective means to investigate the chromospheric magnetic field.
This is because a previous study by \citet[][]{2002ApJ...572..626C} have demonstrated that the opacity of the \Halpha{} line in the upper chromosphere is primarily governed by the ionization degree and radiation field, which remain largely unaffected by local temperature fluctuations but are influenced by mass density \citep[][]{2012ApJ...749..136L}.
Recent work by \citet{2019A&A...631A..33B} further corroborated that the \Halpha{} line maintains opacity even within flaring active regions through synthesizing spectra using 3D rMHD simulations.
Furthermore, \citet{2023ApJ...946...38M}, revealed that the weak field approximation (WFA) over the \Halpha{} line core exhibits a morphology of the LOS magnetic field remarkably similar to inversions of the \CaIR{} line.
This finding underscores the potential of \Halpha{} spectropolarimetry for probing the chromospheric magnetic field.

Despite the potential of the \Halpha{} line as a valuable chromospheric diagnostic tool, its utilization in probing the chromospheric magnetic field remains limited. 
One contributing factor is that despite the dominance of the Zeeman effect in the Stokes~$V$ signal, the line's intensity and polarization profiles are sensitive to the 3D radiation field.
Additionally, in scenarios involving weakly magnetized atmospheres, the Stokes~$Q$~\&~$U$ signals are influenced by atomic polarization \citep{2010MmSAI..81..810S, 2011ApJ...732...80S}.
Thus, it is challenging to model the \Halpha{} line using existing inversion codes that adopt 1.5D plane-parallel geometry.

In this paper, we report the line-of-sight (LOS) magnetic field measurements of an active region inferred from the simultaneous multiline spectropolarimetry in the \CaIRII{} and \Halpha{} lines using state-of-the-art non-LTE inversions and weak field approximation methods.
Recently \citet{2023ApJ...946...38M} have analyzed the diagnostic potential of the \Halpha{} line in probing chromospheric magnetic field. However, their observations were limited to a small pore.
Unlike \citet{2023ApJ...946...38M},  in this work, we present a comprehensive analysis of spectropolarimetric observations of a complex active region consisting of 4 umbrae, 1 penumbra, and a region where the \CaIRII{} line core is in emission, a signature of localized heating.
The \CaIRII{} line has a similar formation to that of the \CaIR{} line, is magnetically sensitive, and can be used as a chromospheric diagnostic \citep{2007ASPC..368..139P, 2007ApJ...670..885P}.

\section{Observations} \label{sect:observations}

\begin{figure*}[htbp]
    \centering
    \includegraphics[width=0.48\textwidth]{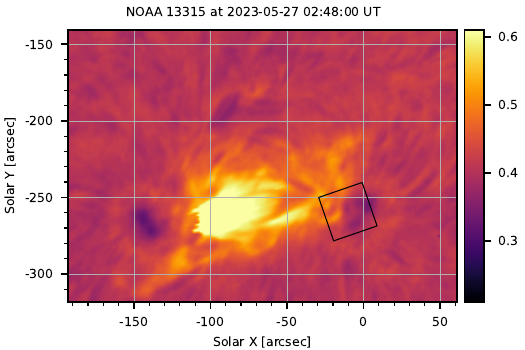}
    \includegraphics[width=0.48\textwidth]{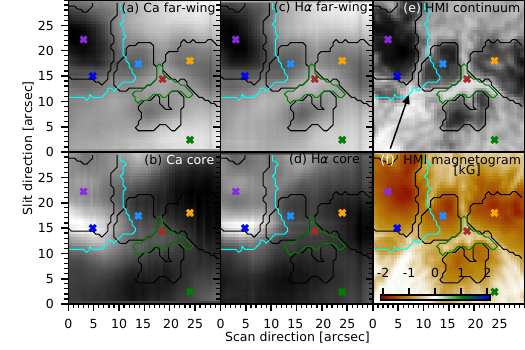}
    \caption{On May 27th, 2023, active region data (NOAA 13315), was recorded. The left panel shows an \Halpha{} filtergram near the core of the \Halpha{} line with FWHM of 0.4~\AA\, recorded using the \Halpha{} telescope at the Kodaikanal Solar Observatory. The observed FOV using the spectropolarimeter at the KTT is marked by the black square and is further shown in right panels. Panels (a) and (b) display images of the \CaIRII{} line: panel (a) shows a far wing image at an offset of $-$0.8~\AA, from the line core, while panel (b) shows the image at the nominal line center. Similarly, panels (c) and (d) show images of the \Halpha{} line: panel (c) presents a far wing image at $-$1.36~\AA, from the line core, and panel (d) shows the image at the nominal line center. Panel (e) is the HMI continuum, and panel (f) is the HMI line-of-sight magnetogram for reference. The arrow shows the direction of the disc center. The black and green color contours represent the umbral and lightbridge regions, respectively. The cyan color contour represents the region showing emission feature in the \Halpha{} filtergram and in the \CaIRII{} line.}
    \label{fig:halphacontext}
\end{figure*}

\begin{figure}[htbp]
    \centering
    \includegraphics[width=0.48\textwidth]{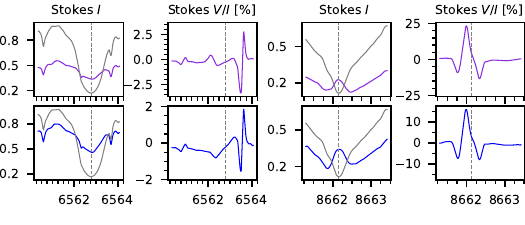}
    \includegraphics[width=0.48\textwidth]{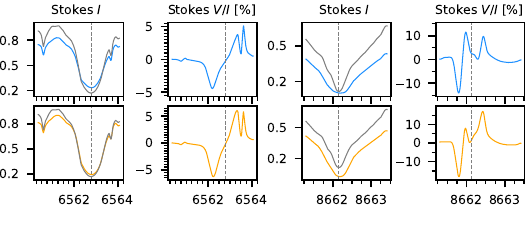}
    \includegraphics[width=0.48\textwidth]{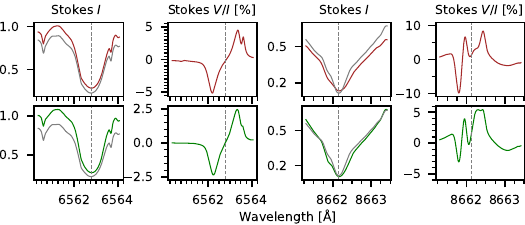}
    \caption{Spectral profiles of a few selected pixels, marked with ``$\times$'' in panels (a)--(f) of Fig.~\ref{fig:halphacontext}. The quiet-Sun profile (solid gray) is also shown for comparison. The gray dashed vertical line shows the position of the nominal line center.}
    \label{fig:fov}
\end{figure}

The observations were made using the polarimeter \citep{2018JATIS...4d8002P} at the Kodaikanal Tower Tunnel (KTT) Telescope \citep{Bappu} at the Kodaikanal Solar Observatory (KoSO) in the \Halpha{} and \CaIRII{} lines simultaneously.
The \FeICaIRIIline{} line, which appears as a blend in the \CaIRII{} spectrum was also recorded.
The spectral sampling of the \Halpha{} and \CaIRII{} lines data were 4.4 m\AA\, per pixel and 3.6 m\AA\, per pixel, respectively.
The pixel scale in the scan direction and slit-direction were binned to 0$\farcs{}$6 per pixel.
The observed field of view (FOV) consists of an active region (NOAA 13315) with multiple sunspots and a light bridge with a viewing angle $\cos \theta = \mu = 0.96$. 
Here, $\theta$ is the angle between the line-of-sight (LOS) direction and the local surface normal.
In each scan step of the raster scan, with step size 0\farcs{}275, we recorded spectropolarimetric data using a 4-stage modulation scheme.
A total of 5 frames were recorded for each modulation state, with each frame having an exposure time of 500~ms.
The spectropolarimetric raster scan was completed over the course of approximately one hour, beginning at 02:14:42 UT and concluding at 03:12:19 UT on May 27th, 2023.
A tip-tilt system \citep{Mathur24} was used during observations.
The data were corrected for dark- and flat-field variations, instrumental polarization, and polarization fringes.
The details are given in the appendix~\ref{sect:reduc}.
The \CaIRII{} data were aligned with the \Halpha{} data by cross-correlating the far wing images of the field-of-view (FOV) at the \CaIRII{} and \Halpha{} lines.
Further, the \Halpha{} data were co-aligned with the Helioseismic and Magnetic Imager (HMI) continuum image.
We also show an \Halpha{} filtergram with FWHM of 0.4~\AA\, recorded using the \Halpha{} telescope at the Kodaikanal Solar Observatory as a context image and describe further in the text. 

An overview of observations is shown in Figure~\ref{fig:halphacontext}.
The panel on the left shows an emission feature over the active region NOAA 13315.
The observed FOV, marked by a black square in the left-panel and shown in panels on the right, covers the emission region partially and the sunspot with multiple structures.
The continuum images for both the \Halpha{} and \CaIRII{} lines exhibit a resemblance to the HMI continuum image (see panels (a), (c), and (e)).
The active region contains multiple structures of sunspot joined by a lightbridge.
The images at the core of the \Halpha{} and the \CaIRII{} lines appear similar (see panels (b) and (d)).
%
%
The active region contains strong fields of negative polarity with a maximum field strength of up to $-$1900~G (see panel (f)).
In both the \Halpha{} and \CaIRII{} line core images, a prominent enhanced emission feature, exhibiting its peak emission about scan direction 5$\arcsec$ and slit position 15$\arcsec$ within the field of view (FOV), is seen.

The quiet-Sun profiles for both the \Halpha{} and \CaIRII{} lines were generated through the averaging of profiles obtained from a select number of pixels within a region situated far from the sunspot from the extended FOV not shown in the right panels of Fig.~\ref{fig:halphacontext}.
The procedure for wavelength calibration, as detailed in appendix~\ref{sect:reduc}, involved comparing the quiet-Sun profile with that of the BASS 2000 atlas.
Additionally, corrections for spectral veil were applied to both the \CaIRII{} and \Halpha{} data.

A few selected spectral profiles corresponding to various dynamics of the FOV, marked in panels (a)-(f) of Fig.~\ref{fig:halphacontext} by ``$\times$'', are shown in Fig.~\ref{fig:fov} using the same color.
Quiet-Sun profile calculated by averaging a few pixels outside the active region is also shown for comparison.

The purple-colored spectral profile belongs to the pixel in the darkest umbral region seen in the far wing images of the \Halpha{} and \CaIRII{} lines.
This pixel exists in the periphery of the emission feature seen in the core of the \Halpha{} and \CaIRII{} lines.
In the Stokes~$I$ profile of the \Halpha{} line, there is a noticeable decrease in intensity within the line wings, a typical characteristic observed in the dark umbral regions.
Moreover, this profile reveals a heightened intensity at its core, in contrast to the quiet-Sun profile, indicating an enhanced emission in the core of the line.
The Stokes~$V/I$ profile of the \Halpha{} line shows an amplitude of 1\%, however, the \ion{Co}{1} blend has an amplitude of 3\%.
In contrast to the \Halpha{} line, for this pixel the Stokes~$I$ profile of the \CaIRII{} line core is seen in emission.
Although the peak of the emission appears un-shifted with respect to the position of the nominal line core, the two minima on either side of the emission peak have different amplitudes, suggesting line-of-sight velocity gradients.
The Stokes~$V/I$ profile of the \CaIRII{} line is not the typical anti-symmetric two-lobed profile.
This variation is due to the presence of the magnetically sensitive \FeICaIRIIline{} line, which appears as a blend blueward to the core of the \CaIRII{} line.
In case of nominal absorption of Stokes~$I$ of \CaIRII{} line, the red lobe of the Stokes~$V/I$ of the \FeICaIRIIline{} line and the blue lobe of the \CaIRII{} line merge, making it challenging to distinguish their amplitude distinctly.
In the present case of the Stokes~$I$ profile in emission, we have observed a distinct three-lobed structure in the Stokes~$V/I$ profile, with the amplitudes of the Stokes~$V/I$ alternating in sign.
The amplitude of the blue and the red lobes are about 10\%, whereas the amplitude of the central lobe is observed to be about 25\%.

The navy-colored spectral profile is associated with a pixel situated near the peak of the enhanced emission feature observed in the images at the core of the \Halpha{} and \CaIRII{} lines. 
This profile, like its purple counterpart, displays a reduction in the intensity of the line wings in both the \Halpha{} and \CaIRII{} Stokes~$I$ profiles, when compared to their corresponding quiet-Sun profiles.
However, the decrease in intensity is less pronounced than in the purple-colored profile.
This lesser reduction is attributed to the pixel's location being outside the dark umbral area, as seen in the far wing images of the \Halpha{} and \CaIRII{} lines.
The line core of the \Halpha{} line's Stokes~$I$ profile exhibits increased intensity compared to the quiet-Sun profile, and the Stokes~$I$ profile of the \CaIRII{} line appears in emission.
The core intensity of both the \Halpha{} and \CaIRII{} lines is higher than in the purple-colored profile.
The Stokes~$V/I$ profile shapes for both lines are analogous to those of the purple-colored profile, yet the amplitude of the Stokes~$V/I$ profile for the \ion{Co}{1} blend and the middle lobe of the \CaIRII{} line is significantly lower than that of the purple colored profile.
This difference in amplitude can be linked to the weaker magnetic field at the navy-colored pixel's location compared to the field strength at the purple-colored pixel's location, as seen in panel (f) of Fig.~\ref{fig:halphacontext}. 
The amplitude of the Stokes~$V/I$ profile of the \Halpha{} line is about 0.9\%.
The amplitude of the blue and red lobes of the \CaIRII{} line is about 8\%, whereas the amplitude of the central lobe is about 15\%.

The cyan- and yellow-colored profiles represent typical profiles in a sunspot umbra.
The Stokes~$I$ profiles of the \Halpha{} line show nominal absorption profiles, with reduced intensity in the line wings and slightly elevated line core.
The Stokes~$I$ profile of the \CaIRII{} line also shows a nominal absorption profile with reduced intensity in the line wings.
The line core intensity of the cyan-colored profile is similar to the quiet-Sun profile, whereas the line-core intenisty of the yellow-colored profile is lesser than the quiet-Sun profile.
The Stokes~$V/I$ profile of the \Halpha{} line is typical with the amplitude of the \ion{Co}{1} blend (3\%) lesser than the amplitude of the \Halpha{} line (7\%).
As stated earlier, the Stokes~$V/I$ profile of the \CaIRII{} line is mixed with the magnetically sensitive \FeICaIRIIline{} line, which appears as a blend.
Consequently, the red lobe of the \FeICaIRIIline{} overlaps with the blue lobe of the \CaIRII{} line.
This overlap is particularly evident in the yellow-colored profile.
The amplitude of the central lobe, influenced by both the \FeICaIRIIline{} and \CaIRII{} lines, is significantly lower than the blue lobe of the \FeICaIRIIline{} line.
This is anticipated due to the contrasting signs of the Stokes~$V/I$ in the red lobe of the \FeICaIRIIline{} line and the blue lobe of the \CaIRII{} line, when both the photosphere and chromosphere exhibit the same magnetic polarity, which in this case is negative.
This effect is less noticeable in the cyan profile.
The amplitude of the blue lobe of the Stokes~$V/I$ profile of the \CaIRII{} line is approximately 15\% and 17\% for the cyan- and yellow-colored profiles, respectively, while the amplitude of the central lobe is about 12\% and 8\%. 
The amplitude of the red lobe is about 10\% and 16\%.

The profiles colored brown and green correspond to pixels in the light bridge and penumbra, respectively. 
Both the \CaIRII{} and \Halpha{} Stokes~$I$ and $V/I$ profiles display typical absorption.
The intensity in the far wings of the \Halpha{} and \CaIRII{} lines is either comparable to or exceeds that of the quiet-Sun profile.
In both profiles, the amplitude of the Stokes~$V/I$ profile for the \ion{Co}{1} blend is smaller than the dark umbral profiles previously discussed.
The amplitude of the Stokes~$V/I$ profile for the green-colored penumbral profile is lower than all other profiles.
For the brown- and green-colored profiles, the amplitude of the Stokes~$V/I$ profile of the \Halpha{} line is approximately 5\% and 2.5\%, respectively.
The amplitude of the blue lobe of the Stokes~$V/I$ profile of the \CaIRII{} line is around 10\% for the brown profile and 3\% for the green profile, while the amplitude of the central lobe is about 7\% for the brown profile and 4\% for the green profile.
The amplitude of the red lobe is about 9\% for the brown profile and 6\% for the green profile.

\label{sect:methods}
\section{Methods} 

\subsection{Weak field approximation}
The magnetic field from the \Halpha{} spectral line was inferred using the weak field approximation (WFA).
Under WFA the Stokes~$V$ is linearly related to \blos{} and ($\partial I / \partial \lambda$) through \citep{2004ASSL..307.....L}
\begin{equation}
\label{eq:wfa}
    V(\lambda) = - \Delta \lambda_{\mathrm{B}}\;\bar{g}\cos \theta\;\frac{\partial I}{\partial \lambda},
\end{equation}

and 

\begin{equation}
    \Delta \lambda_{\mathrm{B}} = 4.67 \times 10^{-13} \lambda_{0}^{2} B,
\end{equation}

where \deltab{} is expressed in \AA\,, $B$ in Gauss, $\bar{g}$ is effective Landé factor, $\theta$ is the inclination of $B$ with respect to the LOS, and $\lambda_{0}$ is the central wavelength of the spectral line (expressed in \AA).

The \blos{} can be derived from Eq.\ref{eq:wfa} using the linear regression formula \citep[e.g.,][]{2009ApJ...700.1391M},

\begin{equation}
    B_{\mathrm{LOS}} = - \frac{\Sigma_{\lambda} \frac{\partial I_{\mathrm{\lambda}}}{\partial \lambda} V(\lambda)}{C \Sigma_{\mathrm{\lambda}} (\frac{\partial I_{\mathrm{\lambda}}}{\partial \lambda})^{2}} \textcolor{red}{,}
\end{equation}
where $C=4.66 \times 10^{-13}\;\bar{g}\;\lambda_{\mathrm{0}}^{2}$.

We have used $\bar{g}=1.048$ following the investigation done by \citet{1994A&A...291..668C}.
We derived two values of \blos{}, one from the line core (\Halpha{}$\pm$0.15~\AA), and over the line wing of the  \Halpha{} line ([$-$1.5~\AA\,, $-$0.15~\AA] and [$+$0.15~\AA\,, $+$1.5~\AA]).
The spectral blends listed in Table~\ref{tab:wfaexclude} were excluded while calculating \blos{} using the WFA, as have done by \citet{2023ApJ...946...38M}, \citet{2022A&A...659A.179J} and \citet{2020JApA...41...10N}.
In this work, we used a spatially-coupled version of WFA developed by \citet{2020A&A...642A.210M}, which imposes spatial coherency in the WFA results.

\begin{table}[htbp]
    \centering
    \begin{tabular}{ccc}
         \hline\hline
         $\lambda_{\mathrm{0}}$ [\AA] & $\Delta\lambda_{\mathrm{0}}$ [\AA] & Line\\
         \hline
         6562.44 & 0.05 & \ion{V}{2}\\
         6563.51 & 0.15 & \ion{Co}{1}\\
         6564.15 & 0.35 & Unknown\\
         \hline
    \end{tabular}
    \caption{The first and second columns define the spectral blends removed before applying the WFA: $\lambda_{\mathrm{0}}\pm\Delta\lambda$. The third column indicates the element of the transition in case it is known.}
    \label{tab:wfaexclude}
\end{table}

\subsection{Non-LTE inversions}

The MPI-parallel STockholm inversion Code \citep[STiC;][]{2019A&A...623A..74D, 2016ApJ...830L..30D}, served as the tool for retrieving the stratification of atmospheric parameters.
STiC builds upon a modified version of the RH radiative transfer code \citep{2001ApJ...557..389U} and employs cubic Bezier solvers to solve the polarized radiative transfer equation \citep{2013ApJ...764...33D}.
In non-LTE, assuming statistical equilibrium, STiC has the capability to simultaneously fit multiple spectral lines.
To address partial re-distribution effects (PRD), it integrates a fast approximation method \citep[for more details][]{2012A&A...543A.109L}.
STiC adopts a plane-parallel geometry for intensity fitting in each pixel, commonly referred to as the 1.5D approximation.
 Furthermore, STiC utilizes an LTE equation-of-state derived from the library functions within the Spectroscopy Made Easy (SME) package code \citep{2017A&A...597A..16P}.
The atmospheric parameters are stratified using the optical depth scale at 5000~\AA\, (500~nm), denoted as \logtau{}.

We have inverted the Stokes~$I$ and $V$ profiles of the \CaIRII{} and \FeICaIRIIline{} lines simultaneously to infer the stratification of temperature (\temp{}), LOS velocity (\vlos{}), microturbulence (\vturb{}) and LOS magnetic field (\blos{}).
We used a 6-level \ion{Ca}{2} atom.
The \ion{Ca}{2}~IR lines were synthesized in complete re-distribution (CRD) approximation.
The atomic parameters of the \FeICaIRIIline{} line were obtained from Kurucz's line lists \citep{2011CaJPh..89..417K} and synthesized under LTE approximation.
We employed $k$-means clustering to categorize the Stokes~$I$ profiles into distinct clusters based on similarity in shape, ensuring profiles with similar shapes were grouped together.
Subsequently, we conducted inversions on the averaged profile within each cluster to infer the stratification of \temp{}, \vlos{}, and \vturb{}. 
The inferred stratification then served as the initial atmospheric guess for determining the atmospheric parameters across the FOV, following a methodology akin to that employed by \citet{2023ApJ...946...38M} and \citet{2022A&A...668A.153M}.
We note here that a total of 6 sets of nodes (not provided in tabular form) in \temp{}, \vlos{}, and \vturb{} for a total of 30 clusters were used during the inversions.
In the second cycle, we inverted the Stokes parameters using nodes only in the \blos{} at \logtau{} = $-$1 and $-$4.5, respectively.
Previous studies have demonstrated that perturbations in the \blos{} within the range of \logtau{} = $-$4 to $-$5 elicit maximum responses in the Stokes~$V$ profiles of the \CaIRII{} line \citep[][]{2016MNRAS.459.3363Q, 2018A&A...619A..63J, 2019ApJ...873..126M}.
In the third cycle, we used nodes in the \temp{}, \vlos{}, \vturb{}, and \blos{} simultaneously to infer consistent values of the atmospheric parameters.
Further, we have applied a median filter spatially over the atmospheric parameters at each cycle and re-ran the particular cycle, which resulted in not only a better fit of the Stokes parameters but also made the variation of the inferred atmospheric parameters spatially smoother.
We have done this two to three times per cycle until we found a satisfactory fit of the Stokes parameters and spatially smooth variation of atmospheric parameters.
Furthermore, we have inverted a few pixels that fail to converge separately, using different node positions and different initialization of atmospheric parameters. However, such pixels were fewer compared to the full FOV.
The quality of fits of our inversions are described in appendex~\ref{sect:quality}.
We set the average velocity in the darkest (left-most umbral substructure in panel (a) of Fig.~\ref{fig:halphacontext}) umbral region in the photosphere (\logtau range of [$-1$, $0$]) to rest for the absolute velocity calibration.

\section{Results} \label{sec:results}

\subsection{Non-LTE inversion results}

\begin{figure}[htbp]
    \centering
    \includegraphics[width=0.5\textwidth]{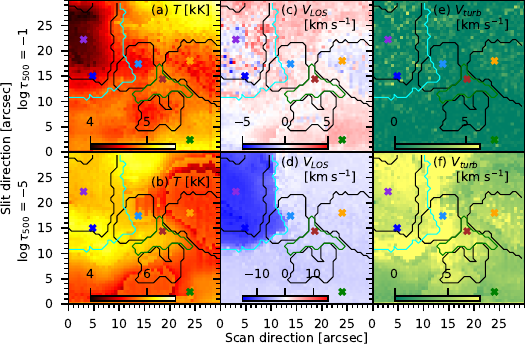}
    \caption{Maps of atmospheric parameters inferred from non-LTE simultaneous multi-line inversions. Panels (a) and (b) depict the maps of \temp{} at $-$1 and $-$5, respectively. Panels (c) and (d) show the maps of \vlos{} at $-$1 and $-$5, respectively. Panels (e) and (f) display the maps of \vturb{} at $-$1 and $-$5, respectively. The contour color scheme is the same as that of in Fig.~\ref{fig:halphacontext}. }
    \label{fig:inversionresult}
\end{figure}

In this section, we will discuss the results from the multi-line inversions of Stokes~$I$ and $V$ profiles of the \CaIRII{} and \FeICaIRIIline{} lines using the STiC inversion code.
In Fig.~\ref{fig:inversionresult}, panels(a)--(f) show maps of \temp{}, \vlos{} and \vturb{} at \logtau{}=$-$1 and $-$5, as indicated in the figure.
%

The \temp{} map at \logtau{} = $-$1 exhibits a morphology akin to the far wing image of the \CaIRII{} line.
In this map, the darkest umbral areas display temperatures around 4~kK, while the majority of the umbra is characterized by a slightly higher \temp{}, approximately 4.5~kK.
The light-bridge area presents a marginally increased \temp{}, about 4.8~kK, in contrast to the penumbral region, which exhibits \temp{} of 5~kK or higher.
Conversely, the \temp{} map at \logtau{} = $-$4.5 shows a morphology comparable to the line core image of the \CaIRII{} line.
Notably, the temperature in the right-most umbral region is lower than in other umbral areas, a characteristic also observed in the line core image of the \CaIRII{} line, where this region appears the darkest.
In the vicinity of the left-most umbral region, where the \CaIRII{} line core is in emission, a higher \temp{}, compared to the rest of the FOV, is seen, around 6.5~kK, a signature of heating.
The lightbridge indicates a \temp{} of approximately 6~kK.

In the \vlos{} map at \logtau{} = $-$1, the umbral region is predominantly static, exhibiting minimal blue or redshifts.
The lightbridge area demonstrates very low downflow velocities, approximately $+$1--2~\kms{}.
The penumbral region, on the other hand, displays downflows in the range of $+$2--3~\kms{}.
In contrast, the \vlos{} map at \logtau{} = $-$4.5 reveals upflows around $-$5~\kms{}.
Notably, in areas where heating is seen. i.e., the \CaIRII{} line is in emission, the upflow velocities exceed $-$10~\kms{}.

There is negligible \vturb{} at \logtau{} = $-1$, whereas at \logtau{} = $-$4.5, there is significant \vturb{} of about 5--8~\kms{}.

\begin{figure}[htbp]
    \centering
    \includegraphics[width=0.5\textwidth]{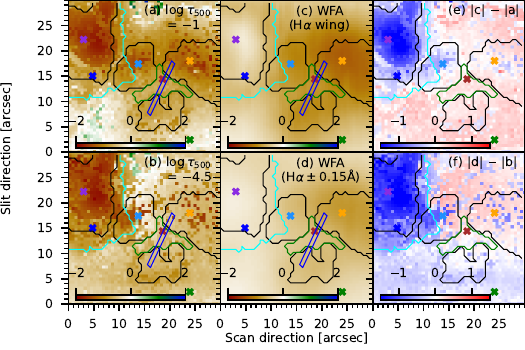}
    \caption{Maps of \blos{} inferred from the inversions and the WFA. Panels (a) and (b) show the maps of \blos{} at \logtau{} = $-$1 and $-$4.5, respectively. Panels (c) and (d) show the maps of \blos{} inferred from the WFA of the \Halpha{} line over the line wings, [$-$1.5~\AA\,, $-$0.15~\AA] and [$+$0.15~\AA\,, $+$1.5~\AA]  and the line core, $\pm0.15$~\AA\,, respectively. Change in the stratification of the $|$\blos{}$|$ is shown in the rightmost two panels (e) and (f) as indicated on each panel. The unit of \blos{} is kG. The contour color scheme is the same as that of in Fig.~\ref{fig:halphacontext}. A blue-colored slit (2-pixel width) is drawn over the lightbridge area which is further discussed in the text.}
    \label{fig:magfield}
\end{figure}

\begin{figure*}[htbp]
    \centering
    \includegraphics[width=0.9\textwidth]{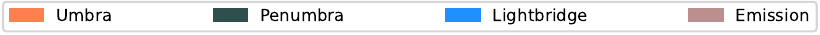}
    \includegraphics[width=0.35\textwidth]{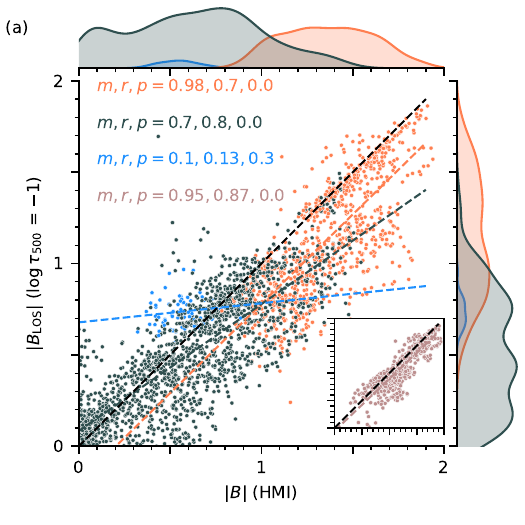}
    \includegraphics[width=0.35\textwidth]{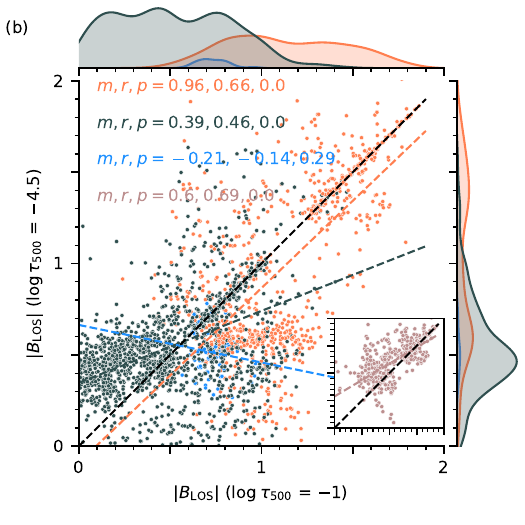}
    \includegraphics[width=0.35\textwidth]{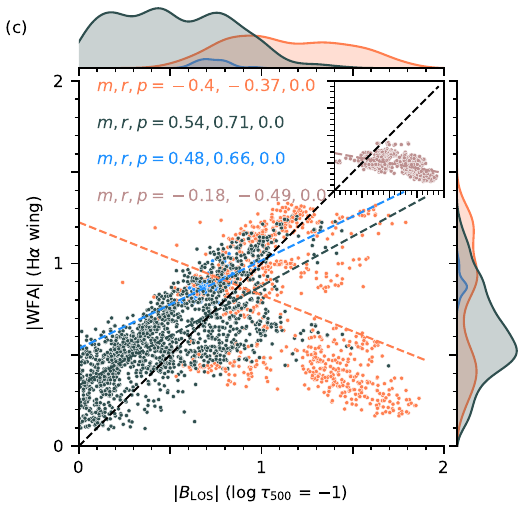}
    \includegraphics[width=0.35\textwidth]{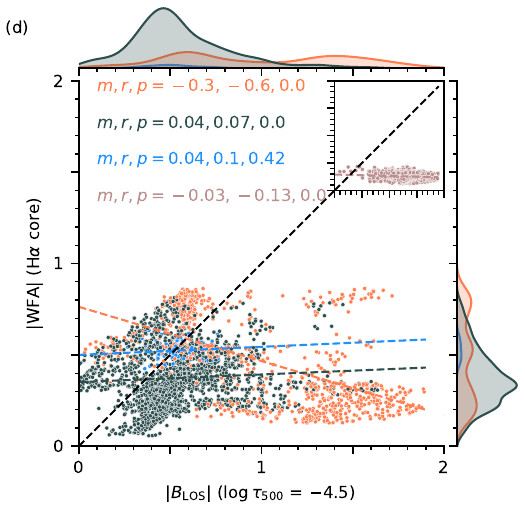}
    \includegraphics[width=0.35\textwidth]{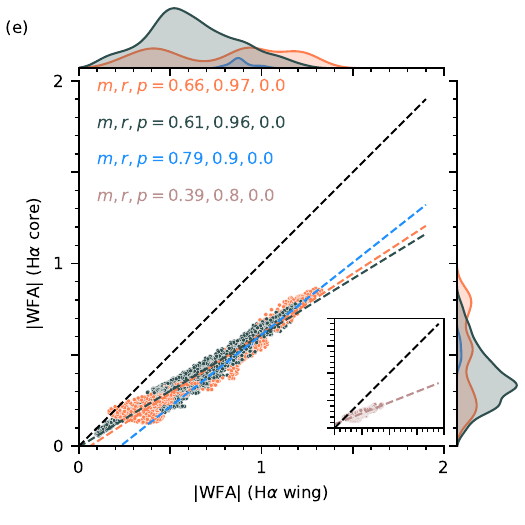}
    \includegraphics[width=0.35\textwidth]{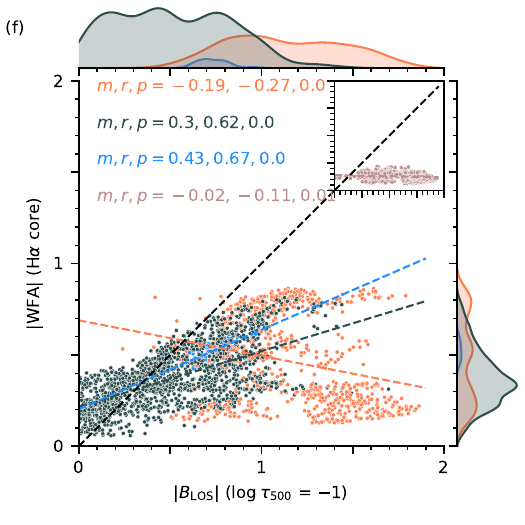}
    \caption{Comparison of the magnitude of \blos{} inferred from inversions and the WFA. Panel (a) shows the comparison between the \blos{} from the HMI magnetogram with that of inferred from inversions at \logtau{} = $-$1. Panel (b) compares between the \blos{} inferred from the inversions at \logtau{} = $-$1 with that of at \logtau{} = $-$4.5. Panel (c) shows the comparison between the \blos{} inferred from inversions at \logtau{} = $-$1 with that of inferred from WFA of the \Halpha{} line over the line wings ([$-$1.5~\AA\,, $-$0.15~\AA] and [$+$0.15~\AA\,, $+$1.5~\AA]). Panel (d) compares between the \blos{} inferred from inversions at \logtau{} = $-$4.5 with that of inferred from WFA of the \Halpha{} line within the spectral range $\pm0.15$~\AA. Panel (e) compares the \blos{} inferred from the WFA over the \Halpha{} line wings with that of inferred from the line core. Panel (f) compares the \blos{} inferred from inversions at \logtau{} = $-$1 with that inferred from WFA over \Halpha{} line core. The unit of the \blos{} is kG. Scatter plots for different regions of the FOV are color-coded and detailed in the legend. Since the emission region contains some part of the leftmost umbral and penumbral region, it is shown separately in the inset plots. The slopes, Pearson correlation coefficient and the p-value of the scatter plot(s) are indicated in the plots by $m$, $r$, and $p$, respectively. The magnetic field inferred using the KTT data at \logtau{} = $-$1 is 0.78 times of HMI magnetogram with $r=0.88$ and $p=0$.}
    \label{fig:magscatter}
\end{figure*}

Panels (a) of Fig.~\ref{fig:magfield} depicts magnetic field map inferred from non-LTE inversions at \logtau{} = $-$1.
Panel (a) of Figure~\ref{fig:magscatter} shows the scatter plots between the magnetic field inferred at \logtau{} = $-$1 and the HMI magnetogram.
The morphological structure of the magnetic field inferred at \logtau{} = $-$1, panel (a), is similar to that of the HMI magnetogram, panel (f) of Fig.~\ref{fig:halphacontext}, with comparable magnetic field strength.
The correlation coefficient between \blos{} at \logtau{} = $-$1 and \blos{} from HMI is of 0.7 and 0.8, for umbral and penumbral regions, respectively, with an overall coefficient of 0.88.
The corresponding slopes are 0.98 and 0.7, with an overall slope of about 0.78.
Notably, there is no correlation between the field strength in lightbridge regions.
It could be because the area of the lightbridge, in terms of the number of pixels, is very small for any meaningful correlation.
It also appears that the field strength from inversions is larger than that of HMI in almost all the pixels in the lightbridge region.
There is a spatial straylight contamination in the data, which is not corrected.
Thus, the stronger fields in the umbral regions near the lightbridge may have some contribution to the fields inferred from the lightbridge region.
Overall, the contribution of the spatial straylight and the seeing-induced spatial smearing could explain the scatter in the magnetic field values and why we could not achieve a one-to-one match with the HMI magnetogram.
The high correlation values found between the field strengths inferred from the data acquired using the polarimeter at the KTT with the standard space-based instruments like HMI suggests that the data acquisition, reduction procedures, and further inversion methods are correctly applied.

Panel (b) of Fig.~\ref{fig:magfield} shows the map of the magnetic field inferred from the inversions at \logtau{} = $-$4.5.
The magnetic field strength and morphology in the umbral regions are similar to the photospheric fields, panel (a).
This can be further verified by examining the panel (b) of Fig.~\ref{fig:magscatter}.
The field strength of the penumbral region at \logtau{} = $-$4.5 is weaker by a factor of 0.39 than that of at \logtau{} = $-$1.
Whereas, the field strength of lightbridge region is negatively correlated by a factor of $-$0.14.
The emission region has a slope of 0.6, however, the field strengths in the emission region are higher at \logtau{} = $-$4.5 than that of at \logtau{} = $-$1.

\begin{figure}[htbp]
    \centering
    \includegraphics[width=0.5\textwidth]{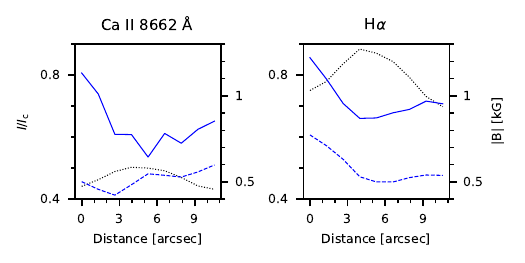}
    \caption{Variation of the magnetic field along the slit drawn in Fig.~\ref{fig:magfield}. The slit width is 2-pixels and all quantities shown are mean within the slit width. The continuum intensity is represented by gray color, and the photospheric and chromospheric magnetic field is represented by solid and dashed curves, respectively.}
    \label{fig:linecut}
\end{figure}

The variation of the mean \blos{} inferred from the inversions and the WFA along a slit (2-pixel width) drawn over the lightbridge in Fig.~\ref{fig:magfield} is shown in Fig.~\ref{fig:linecut}.
The field strength inferred from inversions at \logtau{} = $-$1 in the lightbridge is smaller by about 400~G compared to the umbral region with a stronger field.
Whereas, the field strength inferred from the WFA over \Halpha{} wings is smaller by about 300~G.
The field strength at the \logtau{} = $-$4.5 shows no significant behavior of increase or decrease at the lightbridge compared to the surrounding umbral region.
Whereas, the field strength inferred from the WFA over \Halpha{}$\pm$0.15~\AA\, shows a decrease of about 300~G.

\subsection{Comparison of the \blos{} inferred from the non-LTE inversions and WFA of the \Halpha{} line}

Panels (c) and (d) of Fig.~\ref{fig:magfield} show the maps of the \blos{} inferred from the WFA of the \Halpha{} over the line wing, [$-$1.5~\AA\,, $-$0.15~\AA] and [$+$0.15~\AA\,, $+$1.5~\AA] and the line core, $\pm$0.15~\AA\,, respectively.
This has been done because in the earlier study, \citet{2023ApJ...946...38M}, it was found that the WFA over the line wings and the full \Halpha{} line shows photospheric magnetic morphology, whereas the core of the \Halpha{} line samples the chromospheric magnetic field.
Panels (e) shows the difference in the magnitude between panels (a) and (c) while panel (f) shows the difference in magnitude between panels (b) and (d).
Panel (c) and (d) of Fig.~\ref{fig:magscatter} show the scatter plots between the \blos{} inferred from WFA with that of inversions.
Panel (e) of Fig.~\ref{fig:magscatter} shows the comparison between the \blos{} inferred from the WFA over the \Halpha{} line wing with that inferred from the line core.
Panel (f) compares the \blos{} inferred from the inversions at \logtau{} = $-$1 with that of inferred from WFA over the \Halpha{} line core.

The field strength inferred from the WFA over the line wings of the \Halpha{} line is larger than that of inversions at \logtau{} = $-$1 everywhere in the FOV except in the regions where there is an emission feature in the core of the \CaIRII{} and the \Halpha{} lines and small regions of umbra and penumbra.
However, the scatter plots show, panel (c) of Fig.~\ref{fig:magscatter}, that the slope between the WFA and the inversions is less than 1.
The reason for the increased field strengths is the offset in the linear fit.
The field strength inferred from the WFA over the line wings of the \Halpha{} line in the umbral region, which does not have an emission feature, penumbral region, and the lightbridge has a slope of about 0.5 with respect to that inferred from inversions at \logtau{} = $-$1.
The field strength from the WFA in the region with emission is negatively correlated ($-$0.49) with that of inferred at \logtau{} = $-$1.
Whereas, the full umbra is negatively correlated by a factor of $-$0.37.

Similar to \logtau{} = $-$1, the field strength inferred from the WFA over \Halpha{}$\pm$0.15~\AA\, is smaller than that of inferred from inversions at \logtau{} = $-$4.5 in the region with the emission feature.
Additionally, a large portion of the penumbral region and one of the umbral substructures also show smaller field strengths.
The scatter plot, panel (d) of Fig.~\ref{fig:magscatter}, shows that the field strengths inferred from the WFA over \Halpha{}$\pm$0.15~\AA\, and from inversions at \logtau{} = $-$4.5 are uncorrelated.
However, the majority of the FOV shows smaller field strengths from the WFA than that of inversions.
Interestingly, the morphological structure of the magnetic fields, as inferred from the WFA over the \Halpha{}$\pm$0.15~\AA\, range, exhibits similarities with those inferred from the line wings of the \Halpha{} line, see panels (c) and (d) of Fig.~\ref{fig:magfield}. 
The fields inferred from the \Halpha{} line core are lesser, however correlated with that of inferred from the line wings.
Notably, the emission region shows a lesser slope (0.39) and correlation coefficient (0.8) compared to the whole umbral and penmbral regions, which have a slope of about 0.6 and correlation coefficient of about 0.96, see panel (e) of Fig.~\ref{fig:magscatter}.
The scatter plot of the $|$\blos{}$|$ inferred from the WFA from the \Halpha{} line core appears similar with that of inversions at \logtau{} = $-$4.5 and $-$1, see panels (d) and (f) of Fig.~\ref{fig:magscatter}.
In panel (f), with respect to panel (d), the slope and correlation coefficient is lesser in umbral regions, and higher for penumbral and lightbridge regions.
Interestingly, the slope and correlation coefficient remained same for the emission regions in both the panels (d) and (f).

\section{Discussion}\label{sec:conclusions}

In this paper, we present spectropolarimetric observations of an active region recorded simultaneously in the \Halpha{} and \CaIRII{} lines.
%
%
The stratification of the \blos{} is inferred through multiline inversions of the \CaIRII{}-and-\FeICaIRIIline{} lines and the WFA over the \Halpha{} line.

Consistent with \citet{2023ApJ...946...38M}, we find that the field strengths inferred from the \Halpha{} line core are smaller than that inferred from inversions at \logtau{} = $-$4.5.
Since the opacity of the \Halpha{} line is determined by mass density, \citep{2002ApJ...572..626C, 2012ApJ...749..136L}, and the reasonable assumption that magnetic field expands with height, it is possible that the \Halpha{} line is sampling magnetic field at higher heights than that of the \CaIRII{} line.
It is also to note that in the region with emission feature, the morphology of the magnetic field at \logtau{} = $-$4.5 is remarkably similar to that of at \logtau{} = $-$1, and the field strength is also slightly higher or comparable to that of at \logtau{} = $-$1.
This can be clearly seen in the panel(b) of Fig.~\ref{fig:magfield} and \ref{fig:magscatter}.
It has been found in earlier studies that during events when there is a rise in temperature in the chromosphere, the \ion{Ca}{2} gets ionized to \ion{Ca}{3}, and thus, the \ion{Ca}{2} triplet lines may sample deeper layers of the solar atmosphere \citep{2016ApJ...827..101K, 2018ApJ...860...10K, 2019A&A...631A..33B}. 
In contrast, the field strength inferred in the emission feature from the WFA over the \Halpha{} line core is near constant about 300~G, panel (d) of Fig.~\ref{fig:magscatter}.
The near-constant field strength and the fact that it is much less than the magnetic field inferred from the inversions shows that the field is diffused and the \Halpha{} line is indeed sampling higher heights than the \CaIRII{} line in the region with emission feature.
This is in agreement with the study done by \citet{2019A&A...631A..33B}, that the \Halpha{} line retains opacity in active regions with heating activity.
There is also a good agreement with the field strengths at \logtau{} = $-1$ with that inferred from WFA over \Halpha{} full spectral range, except in the region with emission feature, panel (c) of Fig.~\ref{fig:magscatter}.
In addition, the fields inferred in the emission region from the WFA over \Halpha{}$\pm$0.15~\AA\, and the line wings of the \Halpha{} line are morphologically similar.
Thus, it is possible that in the case of heating events, the full \Halpha{} line becomes sensitive to the chromospheric magnetic field instead of only the line core.
We note here that although the fields inferred from the \Halpha{} line wings and the line core seem correlated, as seen in panel (e) of Fig.~\ref{fig:magscatter}, the similarity of the panel (f) with panel (d) of Fig.~\ref{fig:magscatter}, showing near-zero slopes suggest that the \Halpha{} line core primarily have contribution from the higher atmospheric layers.
Moreover, the emission region is similarly uncorrelated in both the scatterplots.

The WFA over the core of the \Halpha{} line has no correlation with that inferred from inversions at \logtau{} = $-$4.5.
In our earlier study, \citet{2023ApJ...946...38M}, we found that the fields inferred from the \Halpha{} line and the inversions of the \CaIR{} line were correlated.
We provide the following reasons for the lack of correlation in this study.
The field strengths in the earlier study were small, $\leq$600~G, whereas this study focuses on an active region with field strengths of about 2000~G.
It may be possible that the WFA method has a lot of systematic errors in these field ranges.
The validity of the WFA of the \Halpha{} line is not yet quantified using studies based on realistic rMHD simulations.
Additionally, in the earlier study, the spectral profile of pore and surrounding quiet region were simpler compared to this study which has a large sunspot with multiple structures and a large region that shows emission features in the \Halpha{} and \CaIRII{} data, and thus complex spectral profiles.
It may also be possible that there may be systematic errors in the magnetic field inferred in the inversions at \logtau{} = $-$4.5 due to the presence of the \FeICaIRIIline{} line, which appears as a blend very close to the core of the \CaIRII{} line.
There could be a degeneracy between atmospheric parameters that fit the blend as an absorption profile of the \FeICaIRIIline{} line vs an emission feature blueward to the line core of the \CaIRII{} line.
The gradients in the LOS magnetic field and velocity could contribute to the errors retrieved in the \blos{} from the WFA over the \Halpha{} line.
Using semi-empirical models and realistic rMHD simulations that gradients in the LOS velocity and magnetic field introduce errors in the inferred fields from the WFA of the \ion{Ca}{2}~8498 and 8542~\AA\, lines \citep{2018ApJ...866...89C, 2024ApJ...960...26K}.
However, in this study, we suggest that only the effect of magnetic field gradients be a significant factor in the uncertainties in the fields inferred from the WFA of the \Halpha{} line.
This is because we have found that velocity gradients in the \Halpha{} data, calculated using the bisector method and not discussed in the paper, are less than 1~\kms{} and thus not significant.
The $|$\blos{}$|$ inferred from the full spectral range of the \Halpha{} line in all FOV except emission region is weaker by a factor of about 0.6 than that of \blos{} at \logtau{} = $-$1, which is slightly larger than the previous studies (0.42) \citep[][]{1971SoPh...16..384A, 2004ApJ...606.1233B, 2005PASJ...57..235H, 2008ApJ...678..531N, 2023ApJ...946...38M}.

The magnetic field strength within the lightbridge, as inferred from inversions at \logtau{} = $-$1, is observed to be smaller by about 400~G compared to surrounding umbral fields. 
This finding contrasts with earlier studies indicating that lightbridges typically exhibit magnetic field strengths about 1~kG lower than the surrounding umbral fields  \citep{1997ApJ...490..458R, 2003ApJ...589L.117B, 2006A&A...453.1079J, 2008ApJ...672..684R, 2013A&A...560A..84S, 2014A&A...568A..60L, 2017A&A...604A..98J}.
The discrepancy in our measurements may be attributed to factors such as spatial straylight, which remains uncorrected in this study, and seeing-induced smearing.
Additionally, analysis reveals that the magnetic field in the lightbridge at \logtau{} = $-$4.5 does not exhibit a significant difference from the surrounding umbral region, aligning with findings from \citet{ 2017A&A...604A..98J} and \citet{1995A&A...302..543R}, but contradicting the results of \citet{2015SoPh..290.1607S}, who reported lower magnetic field values in the chromosphere at lightbridge locations.
Notably, magnetic field strength inferred from the \Halpha{} line within the lightbridge displays lower values compared to surrounding umbral regions, consistent across both the photosphere (using the full spectral range) and the chromosphere (utilizing the line core).


The heating region displays notable upflows reaching approximately $-$10~\kms{}, potentially indicative of chromospheric evaporation.
Prior observational investigations \citep[for eg.][]{2021A&A...647A.188D, 2021A&A...649A.106Y, 2019A&A...621A..35L, 2008A&A...490..315B, 2002A&A...387..678F} and studies based on rMHD simulations \citep[][]{2019NatAs...3..160C, 2016ApJ...827..101K} have documented instances of chromospheric evaporation.
However, the inferred velocities within the heating region may not be reliable due to the absence of corresponding features in the spectral profiles of the \Halpha{} line.
No asymmetries or line shifts are discernible in the Stokes~$I$ profiles of the \Halpha{} line.
Thus, it is possible that the inversion code utilized these high velocities to fit the observed asymmetry in the two minima on either side of the emission peak of the Stokes~$I$ profiles of the \CaIRII{} line.

\section{Conclusions}

This study presents the stratification of the chromospheric magnetic field using spectropolarimetric observations of an active region in the \CaIRII{} and \Halpha{} lines recorded simultaneously.
In agreement with the \citet{2023ApJ...946...38M}, we found that the magnetic field inferred from the \Halpha{} line core is consistently smaller than that inferred from inversions of the \CaIRII{} line at \logtau{} = $-$4.5.
The magnetic field morphology inferred from the core of the \Halpha{} line resembles that inferred from the full spectral range of the \Halpha{} line in the heating region.
The field strength and morphology inferred in the heating region from the inversions at \logtau{} = $-$4.5 is comparable to that of at \logtau{} = $-$1.
In the heating region, at \logtau{} = $-$4.5, inversions revealed upflows greater than $-$10~\kms{}, whereas signatures of such high upflows were not evident in the \Halpha{} line spectra.
In contrast to the earlier study, \citet{2023ApJ...946...38M}, we found no correlation between the fields inferred from the core of the \Halpha{} line and from inversions at \logtau{} = $-$4.5.
The current study and \citet{2023ApJ...946...38M} highlights the potential of the spectropolarimetric observations of the \Halpha{} line recorded simultaneously with the lines of \ion{Ca}{2} to study the stratification chromospheric magnetic field.
Further spectropolarimetric observations of the \Halpha{} line recorded simultaneously with other chromospheric diagnostics such as \CaIR{} and \HeIIRline{} lines utilizing telescopes with superior spatial and spectral resolution like the Daniel K. Inouye Solar Telescope \citep[DKIST;][]{2020SoPh..295..172R} and the forthcoming European Solar Telescope \citep[EST;][]{2013MmSAI..84..379C} and the National Large Solar Telescope \citep[NLST;][]{2010IAUS..264..499H} are necessary for a comprehensive understanding of the stratification of the chromospheric magnetic field.

\begin{acknowledgments}
This research has made use of the High-Performance Computing (HPC) resources (NOVA cluster) made available by the Computer Center of the Indian Institute of Astrophysics, Bangalore. R.Y. acknowledges support from NSF CAREER award SPVKK1RC2MZ3.

\end{acknowledgments}

\vspace{5mm}
\facilities{KTT\citep{Bappu}}

\software{SunPy\citep{2020ApJ...890...68S},
          NumPy\citep{2020Natur.585..357H},
          matplotlib\citep{2007CSE.....9...90H},
          RH\citep{2001ApJ...557..389U},
          STiC \citep{2019A&A...623A..74D, 2016ApJ...830L..30D}
          }

\bibliography{sample631}{}
\bibliographystyle{aasjournal}

\appendix

\section{Data reduction} \label{sect:reduc}
The data are reduced with standard procedures of bias and flat-fielding.
$I \rightarrow V$ cross-talk is removed by the procedure described in \citet{2022ApJ...930..132J}.
No absolute wavelength calibration is done because of the absence of suitable telluric lines.
Instead, we average a few spatial pixels in the quiescent region outside the active region (quiet-Sun profile) and fit the quiet-Sun profile over the full \CaIRII{} spectral range with the BASS 2000 atlas \citep[][]{2009ASPC..405..397P}.

The procedures of the spectral veil correction, SI intensity calibration, and estimation of spectral Point Spread Function (PSF) for the \CaIRII{}{} and \Halpha{} data are the same as \citet{2023ApJ...946...38M}, which were inspired by \citet{2016A&A...596A...2B}.
The spectral veil and psf is computed by fitting synthesized profiles using FAL-C \citep{1985cdm..proc...67A, 1993ApJ...406..319F} model atmosphere with that of the quiet-Sun profile.
The absolute SI intensity calibration is done by comparing the intensity of the observed predefined continuum wavelength point with the intensity value from psf degraded synthetic profile.
No spectrum tilt was present in the observed data, thus corrections were not necessary.
The full Stokes profiles were filtered using a PCA-based method to improve the signal-to-noise ratio and minimize random fringes in the spectral direction. 
The spectral fringes were further minimized, and spectral profiles were made smoother using a Python implementation of a relevance vector machine based method\footnote{\url{https://github.com/aasensio/rvm}}.

\section{Quality of Fits}\label{sect:quality}
\begin{figure*}[htbp]
    \centering
    \includegraphics[width=0.45\textwidth]{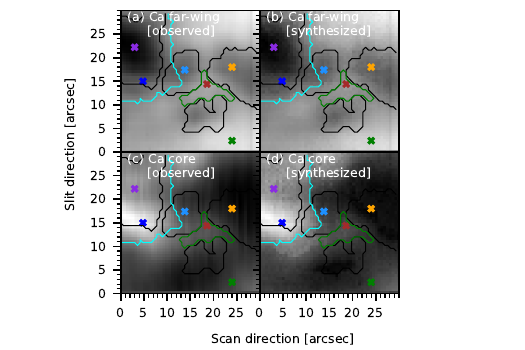}
    \includegraphics[width=0.45\textwidth]{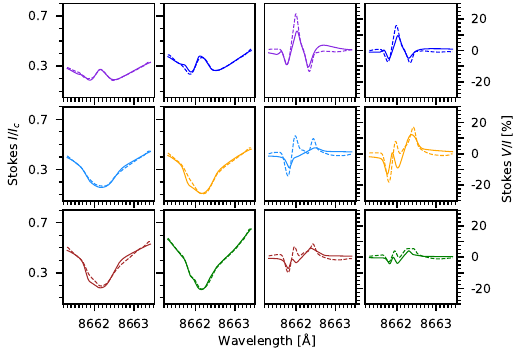}
    \caption{The left panel compares the observed and synthesized images of the FOV near the far-wing and line core wavelength positions of the \CaIRII{} line. The right panel shows the examples of the observed (dotted lines) and synthesized (solid lines) \CaIRII{} Stokes I and V profiles for the pixels marked by ``$\times$'' in Fig.~\ref{fig:halphacontext}.}
    \label{fig:quality}
\end{figure*}

Panels (a) and (b) of Fig.~\ref{fig:quality} compare the observed and synthesized narrowband images at the far wing position of the \CaIRII{} line, whereas panels (c) and (d) compare the narrowband images at the core of the \CaIRII{} line.
The comparison between the synthesized and observed Stokes~$I$ and $V/I$ profiles of a few selected pixels marked by ``$\times$'' in panels (a)--(d) are shown in the right panels with the same color.
The synthesized narrowband images at the far wing and line core positions of the \CaIRII{} line closely resemble those of the observations.
The synthesized Stokes~$I$ profiles of the marked pixels also match very well with observations.
We were able to match the emission feature seen in the purple- and blue-colored profiles as well as a nominal absorption in the remaining profiles.
The synthesized Stokes~$V/I$ profiles show a good match with the observations in the left-most and the right-most lobes.
The amplitude of the central lobe is sometimes lesser than those of the observations.
Overall, we deem the fits to be satisfactory for the analysis done in this paper.

\end{document}